# Hints of the Photonic Nature of the Electromagnetic fields in Classical Electrodynamics and its connection to the electronic charge and vacuum energy density

Vernon Cooray [1], Gerald Cooray [2], Marcos Rubinstein [3] and Farhad Rachidi [4]

[1] Department of Electrical Engineering, Uppsala University, 752 37 Uppsala, Sweden.
[2] Karolinska Institute, Stockholm, Sweden
[3] HEIG-VD, University of Applied Sciences and Arts Western Switzerland, 1401 Yverdon-les-Bains, Switzerland
[4] Electromagnetic Compatibility Laboratory, Swiss Federal Institute of Technology (EPFL), 1015 Lausanne, Switzerland

*Abstract*: The electromagnetic fields of a long dipole working without dispersive and dissipative losses are analyzed in the frequency domains. The dipole produces radiation in bursts of duration $T/2$ where $T$ is the period of oscillation. The parameter studied in this paper is the energy, $U$, dissipated in a single burst of radiation of duration $T/2$. We have studied how $U$ vary as a function of the charge associated with the current in the dipole and the ratio of the length of the dipole and its radius. We have observed a remarkable result when this ratio is equal to the ratio of the radius of the universe to the Bohr radius. Our results, based purely on the classical electrodynamics and general relativity, show that, as the magnitude of the oscillating charge (as defined by the root mean square) reduces to the electronic charge, the energy dissipated in a single burst of radiation reduces to $h\nu$, where $\nu$ is the frequency of oscillation and $h$ is the Planck constant. The importance of this finding is discussed. In particular, the results show that the existence of a minimum free charge in nature, i.e., electronic charge, is a direct consequence of the photonic nature of the electromagnetic fields. Furthermore, the presented findings allow to derive for the first time an expression for the vacuum energy density of the universe in terms of the other fundamental constants in nature, the prediction of which is consistent with experimental observations. This equation, which combines the vacuum energy, electronic charge and mass, speed of light, gravitational constant and Planck constant, creates a link between classical field theories (i.e., classical electrodynamics and general relativity) and quantum mechanics.

## 1. Introduction

Classical electrodynamics is an old subject which has its origins in the 19[th] century with the pioneering work due to James Clerk Maxwell. The subject has been developed and expanded extensively over the years and as it stands today, it is a subject which is fully matured and thoroughly explored. However, as we will show in this paper, there are interesting features hidden within classical electrodynamics which went unnoticed for nearly 160 years by scientists who were more interested either in the basic structure of electrodynamics or its applications in practice.

In analyzing the electromagnetic fields generated by long transmitting dipoles, scientists and engineers utilize electromagnetic field equations pertinent to various charge and current distributions both in time and frequency domain. The expressions for the radiation fields generated by long transmitting dipoles, the subject matter of this paper, are familiar to the scientists and engineers working in the field of antenna theory. However, several interesting features hidden within these electromagnetic field expressions were only recently discovered [1, 2, 3]. The goal of this paper is to improve some of the analysis presented in the above referenced papers and to illustrate these hidden features of the electromagnetic radiation fields.

## 2. Radiation fields and the transport of energy of a long frequency domain dipole

As the radiating system, we will consider a dipole of length $2L$. The relevant geometry is shown in Figure 1. The dipole is located along the z-axis. The dipole is fed by a sinusoidal current of magnitude $I_0$ from the central point (i.e., $z = 0$). If one neglects the dissipation losses, it is a reasonable approximation to assume that the current distribution along the upper half of the dipole arm is given by [4, 5, 6]

$$I(t,z) = I_0 \sin\left\{\frac{2\pi}{\lambda}(L-z)\right\} e^{j\omega t} \quad 0 \leq z \leq L \tag{1}$$

The current distribution in the lower arm of the dipole is given by

$$I(t,z) = I_0 \sin\left\{\frac{2\pi}{\lambda}(L+z)\right\} e^{j\omega t} \quad -L \leq z \leq 0 \tag{2}$$

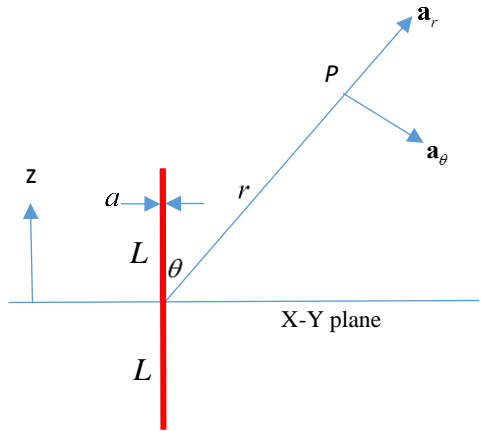

Figure 1: Geometry relevant to the problem under consideration. The dipole is located perpendicular to *x-y* plane with the center of the dipole at the origin. The point of observation P and the relevant distances and angles are marked in the diagram. The unit vector in the direction of increasing polar angle $\theta$ is denoted by $\mathbf{a_\theta}$ and the unit vector in the direction of increasing r is denoted by $\mathbf{a_r}$. The unit vector $\mathbf{a_\phi}$ is directed along $\mathbf{a_r} \times \mathbf{a_\theta}$. The length of the dipole is $2L$ and its radius $a$.

In references [4, 5], one can find the conditions under which this approximation is valid. In general, it is valid if one can neglect all the losses including thermal, dispersive, and dissipative losses.

The electric (in the direction of $\mathbf{a_\theta}$) and the magnetic field (in the direction of $\mathbf{a_\phi}$) at large distances generated by this dipole are given by [4],

$$E_\theta = \frac{j\eta I_0 e^{j\omega t} e^{-jkr}}{2\pi r} \left[ \frac{\cos(kL\cos\theta) - \cos(kL)}{\sin\theta} \right] \quad (3)$$

$$H_\phi = \frac{j I_0 e^{j\omega t} e^{-jkr}}{2\pi r} \left[ \frac{\cos(kL\cos\theta) - \cos(kL)}{\sin\theta} \right] \quad (4)$$

where $k = 2\pi/\lambda$ and $\eta$ is the impedance of free space. The average (over a period of oscillation) Poynting vector associated with these fields is given by

$$\langle \mathbf{S} \rangle = \frac{1}{2} \text{Re}\left[ \mathbf{E} \times \mathbf{H}^* \right] \quad (5)$$

From this, (3), and (4), we obtain

$$\langle \mathbf{S} \rangle = \frac{\eta I_0^2}{8\pi^2 r^2} \left[ \frac{\cos(kL\cos\theta) - \cos(kL)}{\sin\theta} \right]^2 \mathbf{a_r} \quad (6)$$

The total average power dissipated over a single period of oscillation by the dipole is then given by

$$P_{av} = \frac{2 I_0^2}{4\pi\varepsilon_0 c} \int_0^{\pi/2} \left[ \frac{\cos(kL\cos\theta) - \cos(kL)}{\sin\theta} \right]^2 \sin\theta \, d\theta \quad (7)$$

If the oscillating charge associated with the current is given by $q_{osc} = q_0 e^{j2\pi\nu t}$, the above equation can be written as

$$P_{av} = \frac{q^2 (2\pi\nu)^2}{4\pi\varepsilon_0 c} \int_0^{\pi/2} \left[ \frac{\cos(kL\cos\theta) - \cos(kL)}{\sin\theta} \right]^2 \sin\theta \, d\theta \quad (8)$$

In Equation (8) $q$ is the magnitude of the charge as defined by the root mean square value, i.e., $q = q_0/\sqrt{2}$. The integral in (8) can be solved analytically (see reference [4]) resulting in the following expression

$$P_{av} = \frac{q^2 (2\pi\nu)^2}{8\pi\varepsilon_0 c} \left\{ \begin{array}{l} \gamma + \ln(2kL) - C_i(2kL) + \frac{1}{2}\sin(2kL)\left[ S_i(4kL) - 2S_i(2kL) \right] \\ + \frac{1}{2}\cos(2kL)\left[ \gamma + \ln(kL) + C_i(4kL) - 2C_i(2kL) \right] \end{array} \right\} \quad (9)$$

where $C_i$ is the cosine integral, $S_i$ is the sine integral and $\gamma$ is Euler's constant. Observe first that the power generated by the dipole consists of bursts of duration $T/2$, where $T$ is the period of oscillation. The energy dissipated by a single burst of energy $U$ of duration $T/2$, is given by

$$U = \frac{q^2 \pi v}{4\varepsilon_0 c} \left\{ \begin{array}{l} \gamma + \ln(2kL) - C_i(2kL) + \frac{1}{2}\sin(2kL)\left[S_i(4kL) - 2S_i(2kL)\right] \\ + \frac{1}{2}\cos(2kL)\left[\gamma + \ln(kL) + C_i(4kL) - 2C_i(2kL)\right] \end{array} \right\} \quad (10)$$

One can observe from this equation that for large values of $kL$, the value of $U$ oscillates rapidly with $kL$. The upper and lower bounds of $U$ occur when $2kL = n\pi$ and $2kL = m\pi$ where $n$ and $m$ are even and odd integers (i.e., when $\cos(kL) = 1$ or $\cos(kL) = -1$). The median value of $U$ is given by

$$U_{med} = \frac{q^2 \pi v}{4\varepsilon_0 c} \{\gamma + \ln(2kL) - C_i(2kL)\} \quad (11)$$

Note that for large values of $kL$, the cosine integral varies as $\cos(2kL)/(2kL)^2$ and it can be neglected with respect to other terms. Thus, for large values of $kL$, the expression for the median energy reduces to

$$U_{med} = \frac{q^2 \pi v}{4\varepsilon_0 c} \{\gamma + \ln(2kL)\} \quad (12)$$

## 3. Absolute maximum value of the median energy as a function of the oscillating charge

The equations derived in the previous section are valid provided that $\lambda \geq a$, where $a$ is the radius of the dipole. Using this fact, one can see from Equation (12) that for a given dipole length, the maximum value of energy is reached for wavelengths comparable to the radius of the dipole $a$. Thus, the maximum energy for a given charge is obtained for the largest possible value of the ratio $L/a$. Let us consider the natural limits imposed on this ratio by nature. The smallest possible radius of a dipole that can exist in nature is equal to the Bohr radius $a_0$, i.e., the atomic dimensions. The largest possible value of the dipole length $L$ that one can have in nature is equal to the radius of the universe. Since the universe is expanding, the radius of the universe changes with time. However, according to the current understanding, the radius of the universe (the region where events are in causal contact) becomes constant at some future epoch and this value is given by the steady state value of the Hubble radius [7, 8, 9,10]. Let us denote this radius by $R_\infty$. This radius is given by, $R_\infty = c^2 \sqrt{3/8\pi G \langle \rho_\Lambda \rangle}$ where $\langle \rho_\Lambda \rangle$ is the vacuum energy density, $c$ is the speed of light and $G$ is the gravitational constant [7]. Thus, the upper limit of the median energy dissipated within a single power burst of duration $T/2$ for a given charge is given by,

$$\langle U_{med} \rangle_{max} = \frac{q^2 \pi v}{4\varepsilon_0 c} \{\gamma + \ln(4\pi R_\infty / a_0)\} \quad (13)$$

Figure 2 shows a plot of $\langle U_{med} \rangle_{max}$ as a function of the root mean square value of the oscillating charge. This equation predicts that when $q = e$, where $e$ is the electronic charge, $\langle U_{med} \rangle_{max} \approx h\nu$ (within 0.06% if we use the measured value of $5.356 \times 10^{-10}$ J/m$^3$ for $\langle \rho_\Lambda \rangle$) where $h$ is the Planck constant. Since $\langle U_{med} \rangle_{max}$ is the maximum energy that can be radiated within a single burst of power for any given charge $q$, the results can be summarized by the mathematical statement

$$U \geq h\nu \Rightarrow q \geq e \quad (14)$$

where $U$ is the energy radiated in a single power burst of duration $T/2$, $h$ is the Planck constant, $\nu$ is the frequency of oscillation and $q$ is the root mean square value of the oscillating charge. First, observe that the right-hand side of the mathematical statement is valid because according to the experimental data the smallest free charge that can exist in nature is the electronic charge. Given this fact, the smallest value of the charge that oscillates in the dipole has to be either equal to or larger than the electronic charge. Second, since we have considered the maximum ratio $L/a$ ever possible by a dipole, this mathematical statement is universal and it is satisfied by any dipole. Third, it is important to point out that the reverse of this mathematical statement, i.e., $q \geq e \Rightarrow U \geq h\nu$ is not necessarily true. For example, even if the charge is larger than the electronic charge, by decreasing the size of the dipole one can make the energy be sufficiently low for the relation $U < h\nu$ to be satisfied. Finally, it is important to point out that the derivation is purely based on classical electrodynamics and the appearance of the Planck constant in the expression is purely due to the use of atomic units to describe the energy.

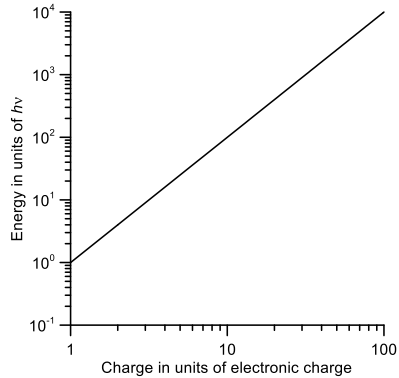

Figure 2: The absolute value of the median energy transported by a single burst of radiation, $\langle U_m \rangle_{max}$, as a function of the root mean square value of charge as given by Equation (13).

## 4. Discussion and the significance of the results

Note that the ultimate dipole that we have considered here is only a theoretical possibility and it could never be realized in practice. Moreover, in deriving the results presented here we have assumed ideal conditions without losses, which also cannot be realized in practice. In this respect, we have to treat the problem we have analyzed as a 'Gedanken' experiment. The goal of such an experiment here is to extract the ultimate limits of a given theory.

One may question whether the mathematical statement derived here (i.e., $U \geq h\nu \Rightarrow q \geq e$) will be disturbed by the losses that will always be present in the actual dipoles. The effect of losses is to reduce the radiated energy associated with the oscillating charge and, for this reason, the derived mathematical statement will still be valid even in the presence of losses. Note also that the presence of Planck constant in the inequality derived here might incorrectly give the impression that quantum mechanics is involved in this derivation. This is not the case. The derivation is purely classical without any involvement of quantum mechanics. It is only our choice of atomic units for the energy that introduced the Planck constant into the equations.

Earlier we have pointed out that based on the current experimental knowledge, the right hand side of the inequality is valid. Now, the left-hand side of the inequality states that $U \geq h\nu$. The validity of this inequality cannot be assertained within the confines of the classical electrodynamics. To do that we have to appeal to quantum mechanics. First, let us assertain the general validity of the inequality and, based on that, discuss the significance of the results.

Recall that $U$ is the energy dissipated within the time $T/2$, the duration of a single burst of radiation. According to quantum mechanical interpretation, the electromagnetic radiation consists of photons. If any electromagnetic energy is dissipated within the time duration $T/2$, then at least one or more photons should be released by the radiator during this time interval. Since the energy of a photon cannot be smaller than $h\nu$, the energy released during the time period $T/2$ cannot be smaller than $h\nu$. In other words, the quantum mechanical nature of electromagnetic radiation gives rise to the inequality $U \geq h\nu$.

The discussion above shows that not only the right-hand portion of the mathematical statement given by Equation (14) but also the left-hand portion of the mathematical statement is also correct. This finding suggests that the condition $q \geq e$ is a consequence of the quantum nature of the electromagnetic fields. The result is significant in two ways. First, knowing the fact that the electronic charge is the minimum charge, the classical electrodynamics itself could have come up with one of the most fundamental concepts in quantum mechanics: the quantum nature of electromagnetic radiation. Second, the fact that the minimum free charge that exists in nature is equal to the electronic charge is shown for the first time to be a direct consequence of the photonic nature of the electromagnetic fields.

Consider the expression for the energy as given by Equation (13). Substituting this in Equation (14) we obtain

$$\frac{q^2 \pi \nu}{4\varepsilon_0 c}\{\gamma + \ln(4\pi R_\infty / a_0)\} \geq h\nu \Rightarrow q \geq e \quad (15)$$

Considering the equal signs of inequalities on both sides we obtain the expression

$$\frac{e^2 \pi}{4\varepsilon_0 c}\{\gamma + \ln(4\pi R_\infty / a_0)\} = h \quad (16)$$

In the above equation, $R_\infty$ is the steady-state value of the Hubble radius. Observing that $R_\infty = c^2\sqrt{3/8\pi G \langle\rho_\Lambda\rangle}$ where $\langle\rho_\Lambda\rangle$ is the vacuum energy density [7], the above equation can be written as

$$\frac{e^2\pi}{4\varepsilon_0 c}\left\{\gamma + \ln(4\pi c^2 \sqrt{3/8\pi G \langle\rho_\Lambda\rangle}/a_0)\right\} = h \quad (17)$$

This equation connects the electronic charge to the vacuum energy density through other constants of nature. By making the vacuum energy density the subject, we can write

$$\langle\rho_\Lambda\rangle = \left\{\frac{24\pi^3 m_e^2 c^6 e^{2\gamma}}{Gh^2}\alpha^2\right\} e^{-\frac{4}{\pi\alpha}} \quad (18)$$

Note that we have replaced $a_0$ by $h/2\pi m_e c\alpha$ where $m_e$ is the rest mass of the electron and $\alpha$ is the fine structure constant which is equal to $e^2/2\varepsilon_0 ch$. Equation (18) can be written as

$$\langle\rho_\Lambda\rangle = \left\{\frac{m_e c^2}{(2\pi\lambdabar_e)(\pi l_p^2)}\right\} f(\alpha) \quad (19)$$

with

$$f(\alpha) = 12\pi^2 \alpha^2 e^{-\frac{4}{\pi\alpha}+2\gamma} \quad (20)$$

In the above equation, $\lambdabar_e$ is the Compton radius of the electron and $l_p$ is the Planck length. Observe that the quantity inside the curly bracket is equal to the energy density that will result if the mass of the electron is confined to a ring of radius equal to the Compton radius of the electron with a cross sectional thickness equal to the Plank length. We would not like to give much physical significance to the quantity inside the bracket, but it is of interest to mention that starting from the pioneering work of Compton [11], there are many studies that attempt to model the electron as a very thin ring with a radius equal to the Compton radius. We may also add that, to the best of our knowledge, the cross-sectional thickness of the ring has not been equated to be the Planck length. The energy density term inside the curly bracket is huge but it is brought down to a small value by the action of the fine structure constant through the function $f(\alpha)$. Equations (18) and (19) predict $\langle\rho_\Lambda\rangle = 4.3\times10^{-10}$ J/m$^3$ which is in agreement with the current experimentally observed value of $5.356\times10^{-10}$ J/m$^3$ [8, 9, 10]. This is a significant result given the current state of the theoretical estimations of vacuum energy density. This equation, which combines together the vacuum energy, electronic charge and mass, speed of light, gravitational constant and Planck constant, creates a link between classical field theories (i.e., classical electrodynamics and general relativity) and quantum mechanics.

After the observation of the accelerated expansion of the universe, scientists have invoked vacuum energy as a possible explanation for this rapid expansion. However, several attempts to theoretically estimate the vacuum energy density gave rise to values which are about $10^{120}$ times larger than the observed value [12]. Modern research work led to a value which is about $10^{60}$ larger than the observed value. This discrepancy, known as the vacuum energy catastrophe, is named as one of the worst predictions in physics. The results presented here provide an equation that predicts the value of the vacuum energy density in terms of other fundamental constants to a reasonable accuracy.

Observe that the vacuum energy density is strongly connected to the fine-structure constant which determines the strength of the electromagnetic force. According to Equation (18) or (19), the vacuum energy density decreases with decreasing fine-structure constant and vice versa. To the best of our knowledge, this is the first time that a relationship is derived for the vacuum energy in terms of the other fundamental constants of nature. This expression for the vacuum energy density, which gives a reasonable estimate in agreement with experiment is based on classical field theory (general relativity and classical electrodynamics) in combination with the quantization of the electromagnetic energy. The narrative presented here is somewhat similar to the estimation of the Bekenstein-Hawking entropy of a black hole based on general relativity and thermodynamics (which similarly is overestimated using standard quantum field theories) [13, 14].

It is interesting to note that based on the Large Number Hypothesis of Dirac [15], it had been speculated based on numerical coincidences that the fine-structure constant, $\alpha$, changes as the inverse of the logarithm of the age of the universe expressed in Planck units [16]. Consider Equation (16). It can be written as

$$\alpha \approx \frac{1}{\left\{\ln(4\pi \langle R_\infty \rangle / a_0)^{\pi/2}\right\}} \quad (21)$$

After substituting for the parameters inside the logarithmic term we obtain

$$\alpha = \frac{1}{\ln(1.326 \times 10^{59})} \quad (22)$$

The number inside the logarithmic term in the above equation is approximately equal to the current age of the universe in Planck units of time. This, therefore, provides a possible explanation for the numerical coincidence $\alpha \sim 1/\ln(t)$ where $t$ is the current age of the universe in Planck units. This point illustrates that extracting information from numerical coincidences in nature should be done with caution.

While some physicists are trying to understand and resolve the vacuum energy catastrophe, others are suggesting that, similar to the electronic charge or the fine-structure constant, one should treat the vacuum energy density as a natural constant whose value is fixed by experimental observations. Some physicists raise the philosophical question: What is the reason for the vacuum energy density to have an extremely small value instead of zero? Equation (18) or (19) provides an answer to this question. This equation shows that the magnitude of the vacuum energy density is decided by the value of the fine-structure constant. In this respect the vacuum energy density could be treated as a dependent parameter. The equation also shows that a zero vacuum energy density will arise only in a universe which does not contain the electromagnetic force.

Finally, we would like to point out that we do not attempt to ascertain that the classical electrodynamics is a complete theory. We understand that classical electrodynamics is an approximate theory and the correct theory is the quantum electrodynamics. Quantum electrodynamics is needed when one analyses problems related to photon–photon scattering and many other phenomena related to photons or virtual photons, and quantum entanglement. However, according to Jacksson [5], classical electrodynamics may in many cases describe the average behavior of the system even in cases where the quantum electrodynamics is needed to describe the individual events. It is important to point out that the electromagnetic fields, whose average behavior is described accurately by classical electromagnetics, are, at a fundamental level, photonic in nature. Thus, it is not surprising that, when driven to its extreme limits, classical electrodynamics could reveal this quantum nature of the electromagnetic fields. This happened to be the case with the problem we have analyzed in this paper.

Note also that the problem we have analyzed here is a classical problem where classical electrodynamics is applicable. Of course, we have used the Bohr radius as the radius of the dipole in our analysis. From a classical point of view, there is no restriction to the use of Bohr radius as the radius of the conductor because from a classical point of view there is no restriction to the radius of the dipole. For example, classical electrodynamics have been used to study the radiating systems with infinitely small radii, namely, thin wire antennas [5]. It is also important to point out that our choice of Bohr radius as the radius of the dipole places the high frequency limit of our system well below the frequencies where pair-production takes place.

The interesting point is that the classical result shows that the existence of the electronic charge is a direct consequence of the photonic nature of the electromagnetic fields. Observe, however, that the predicted value of the energy when $q = e$ is not exactly equal to the energy of a photon but it is extremely close to it. This slight discrepancy is probably caused by our use of classical electrodynamics. One question that one can raise in this respect is the following: Had one used the quantum electrodynamics including the corrections to the Maxwell's equations as derived by Euler and Heisenberg [17] to analyze this problem, which of course is beyond the scope of this paper, would the results obtained be different to the ones obtained here? Indeed, the quantum electrodynamics, the foundation of which is the quantization of the electromagnetic field, would have confirmed and made these predictions even more accurate because the predictions of quantum electrodynamics would adhere to the foundation on which it is built.

## 4. Conclusion

The minimum energy that can be generated by an electromagnetic radiator with a length equal to the ultimate radius of the universe is analyzed. Our study shows that:

(i) Knowing the fact that the electronic charge is the minimum charge, the classical electrodynamics itself could have come up with the most fundamental concept in quantum mechanics: the quantum nature of electromagnetic radiation, and,

(ii) the fact that the minimum free charge that exists in nature is equal to the electronic charge is shown for the first time to be a direct consequence of the photonic nature of the electromagnetic fields.

From this analysis, we have obtained an expression for the vacuum energy density (in effect Einstein's cosmological constant) using gravitational and electromagnetic constants and the quantum nature of energy. The estimated value for the vacuum energy density using the derived expression agrees with experimental observations.

The physical connection we obtained between vacuum energy density and the fine-structure constant (or electronic charge) could be an indication that the nature of vacuum energy, at least to a large extent, is electromagnetic.